\documentclass[aps,nofootinbib,preprintnumbers,twocolumn,superscriptaddress,prd]{revtex4-1}

\usepackage{amsmath}
\usepackage{amssymb}
\usepackage{slashed}
\usepackage{epsfig}
\usepackage{bbm,bm}
\usepackage{hyperref}
\usepackage{color}
\usepackage{comment}
\usepackage{amsfonts}
\usepackage{dsfont}

\makeatletter
\@addtoreset{equation}{section}

\makeatother

\def\bea{\begin{eqnarray}} 
\def\eea{\end{eqnarray}}
\def\be{\begin{equation}} 
\def\ee{\end{equation}} 
\def\ba{\begin{array}}
\def\ea{\end{array}} 
\def\nn{\nonumber}

\def\be{\begin{equation}}
\def\ee{\end{equation}}
\def\bea{\begin{eqnarray}}
\def\eea{\end{eqnarray}}

\def\nn{\nonumber}

\let\oldtitle\title
\renewcommand{\title}[1]{\oldtitle{\color{blue}{#1}}}

\let\oldeqref\eqref
\let\oldcite\cite
\renewcommand{\eqref}[1]{{\color{blue}\oldeqref{#1}}}
\renewcommand{\cite}[1]{{\color{blue}\oldcite{#1}}}

\makeatletter
\let\reftagform@=\tagform@
\def\tagform@#1{\maketag@@@{\ignorespaces\textcolor{blue}{(\ignorespaces #1 \unskip\@@italiccorr \ignorespaces)\ignorespaces}}}
\makeatother

\makeatletter
\renewcommand{\p@subsection}{}
\renewcommand{\p@subsubsection}{}
\makeatother

\usepackage{mathpazo}
\usepackage{amsmath}

\begin{document}

\title{
New universality class in three dimensions:
The critical Blume-Capel model
}

\author{A.\ Codello}
\email{codello@cp3-origins.net}
\affiliation{CP$^3$-Origins, 
University of Southern Denmark,
Campusvej 55, 5230 Odense M, Denmark}
\affiliation{INFN - Sezione di Bologna, via Irnerio 46, 40126 Bologna, Italy}

\author{M.\ Safari}
\email{safari@bo.infn.it}
\affiliation{INFN - Sezione di Bologna, via Irnerio 46, 40126 Bologna, Italy}
\affiliation{
Dipartimento di Fisica e Astronomia,
via Irnerio 46, 40126 Bologna, Italy}

\author{G.\ P.\ Vacca}
\email{vacca@bo.infn.it}
\affiliation{INFN - Sezione di Bologna, via Irnerio 46, 40126 Bologna, Italy}

\author{O.\ Zanusso}
\email{omar.zanusso@uni-jena.de}
\affiliation{
Theoretisch-Physikalisches Institut, Friedrich-Schiller-Universit\"{a}t Jena,
Max-Wien-Platz 1, 07743 Jena, Germany}
\affiliation{INFN - Sezione di Bologna, via Irnerio 46, 40126 Bologna, Italy}

\begin{abstract}
We study the {\tt Blume-Capel} universality class in $d=\frac{10}{3}-\epsilon$
dimensions. The RG
flow is extracted by looking at poles in fractional
dimension of three loop diagrams using $\overline{\rm MS}$. The theory is the only nontrivial
universality class which admits an expansion to three dimensions with
$\epsilon=\frac{1}{3}<1$. We compute the relevant scaling exponents
and estimate some of the OPE coefficients to the leading order.
Our findings agree with and complement CFT results.
Finally we discuss a family of nonunitary multicritical models which includes the {\tt Lee-Yang} and {\tt Blume-Capel}
classes as special cases.
\end{abstract}

\maketitle


\section{Introduction}

The universal behavior of macroscopic systems has long been attracting the interest of the scientific community
because it shows unexpected connections among several
different areas of physics and draws interdisciplinary connections with other quantitative sciences.
It is well known that systems undergoing a second order phase transition exhibit a diverging correlation length at the critical point,
which typically signals the separation among two or more macroscopically distinct phases.
In fact, close to a second order critical point
the system \emph{forgets} the details of its microscopic interactions
because of the large correlation length, and therefore very different microscopic models
might exhibit the same macroscopic behavior. Such models are said to constitute a \emph{universality class}.

The most famous second order phase transition is perhaps
the one observed in ferromagnetic systems which demonstrate a separation between ordered and disordered magnetic phases,
and which could be described by the critical Ising model with nearest neighbor interactions among microscopic spins.
Interestingly the same
critical properties are observed close to the critical point
of the liquid-vapor transition in the
phase diagram of water.
The two physical systems therefore belong to the same universality class,
which is known to be described by the $\phi^4$ model with scalar order parameter $\phi$ \cite{Pelissetto:2000ek}.

The study of universality classes systematizes our understanding
of long range interactions in critical systems.
On the one hand, it is often possible to identify the order parameter $\phi$
and study the Ginzburg-Landau description of the critical system in terms of its free energy
and its renormalization group (RG).
On the other hand, it has been observed that the scale invariance of a critical point
is often promoted to full conformal invariance.
Solid investigations thus interpolate various methods of field theory, 
including perturbation theory, RG and conformal field theory (CFT) methods.

Even more interestingly, Nature is not promiscuous
in that it seems to provide us with a comparatively small number of universality classes
in three dimensions, making the discovery of any new one
even more interesting.
In fact, while in two dimensions there is a countable family of critical models,
including the notable examples of the CFT minimal models ${\cal M}_{p,q}$ \cite{Belavin:1984vu},
in three dimensions there is, in comparison, a scarcity.
This is especially true in the absence of global symmetries and for a single scalar order parameter $\phi$,
in which case there might be only three such models:
the {\tt Ising} universality
class,\footnote{We use {\tt typewriter} font to denote universality classes.} with upper critical dimension $d_c=4$ \cite{Brezin:1972fc}
the {\tt Lee-Yang} universality class with $d_c=6$ \cite{Fisher:1978pf}, and the {\tt Blume-Capel} universality class with $d_c=\frac{10}{3}$ which 
is the object of this paper.\footnote{The {\tt Tricritical} Ising universality class has $d_c=3$ and therefore is Gaussian in $d=3$, while
the {\tt Blume-Capel} universality class is a tricritical generalization of the {\tt Lee-Yang} class \cite{Nicoll:1974zz}
and as such we refer to it following \cite{vonGehlen:1994rp,Zambelli:2016cbw}.}
These three models are all believed to be CFTs at criticality \cite{ElShowk:2012ht,Gliozzi:2016ysv,Alday:2016njk,Rychkov:2015naa,Basu:2015gpa,Nii:2016lpa,Codello:2017qek}.

The {\tt Blume-Capel} universality class has some rather interesting properties:
its upper critical dimension is a rational number slightly above three,
meaning that the model offers a new nontrivial critical point in three dimensions.
In an $\epsilon$-expansion with $d=\frac{10}{3}-\epsilon$ it is sufficient to set $\epsilon = \frac{1}{3}$
to estimate the physically interesting case $d=3$. The $\epsilon$-expansion is thus expected to be better defined
and able to give more precise estimates if compared with the other two nontrivial three dimensional universality classes.
Another property of this universality class is that the {\it leading} perturbative RG flow
must be obtained by looking at $\frac{1}{\epsilon}$ poles
of {\it three loop} diagrams.

In this paper we compute for the first time the leading order corrections
in the $\epsilon$-expansion to the spectrum and the operator product expansion (OPE) coefficients of the {\tt Blume-Capel} class.
For this purpose, we use the powerful functional perturbative RG methods recently developed \cite{ODwyer:2007brp,Codello:2017hhh}.
A preliminary analysis of the conformal data has been performed in \cite{Codello:2017qek} with CFT methods,
but up to now only with RG methods it
is possible to obtain the critical coupling at the fixed point $g(\epsilon)$, which is the gateway for numerical estimates of critical quantities in $d=3$.

We expect that the new critical point can be observed either theoretically in computer simulations,
or experimentally in opportunely tuned
systems such as the atomic mixtures described by the microscopic Blume-Capel model \cite{BC}
which has enough degrees of freedom to exhibit the tricritical phase \cite{vonGehlen:1994rp}.
The critical point might also be relevant in the understanding of the full analytic structure of the partition function of the
tricritical Ising model as a function of the magnetic field \cite{Mossa:2007fx}.

Finally, we complement the analysis by covering a family of multicritical nonunitary models
which includes the {\tt Lee-Yang} and {\tt Blume-Capel} classes as the first two special examples.
All models besides the first two have upper critical dimension smaller than three, and thus are physically
interesting in two dimensions, where they are expected
to correspond to a nonunitary subset of the CFT minimal models ${\cal M}_{p,q}$ \cite{vonGehlen:1994rp,Zambelli:2016cbw,Castro-Alvaredo:2017udm}.

\section{Beta functionals}

\begin{figure}[tb]
\begin{center}
\includegraphics[width=0.3\textwidth]{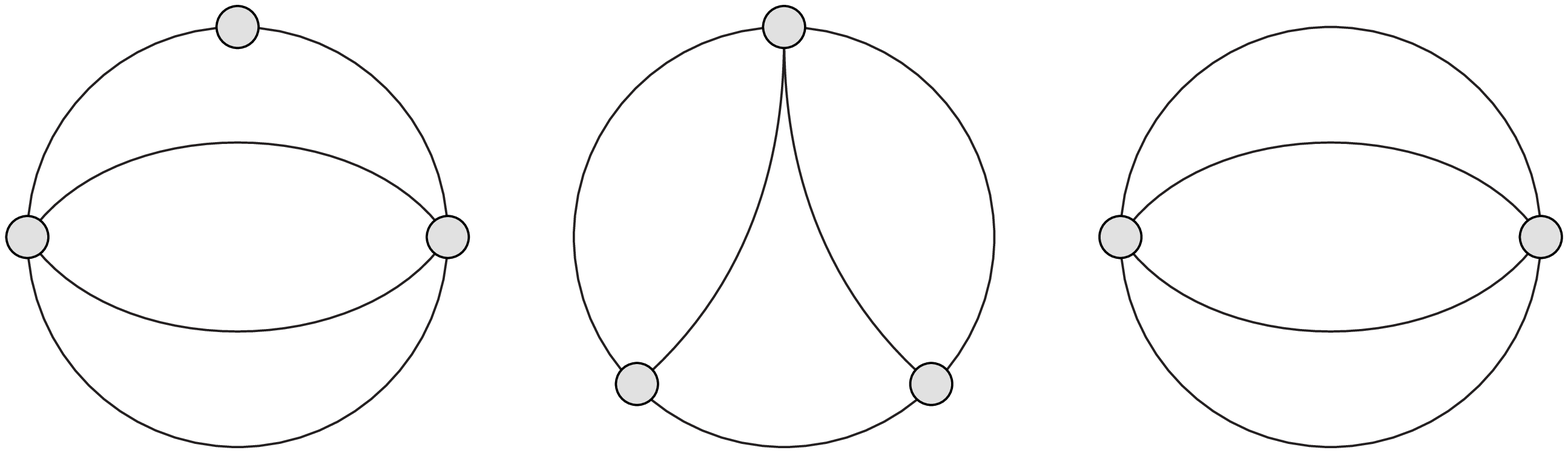}
\caption{
Diagrams responsible for $\beta_V$ and $\beta_Z$.
The lines represent massless scalar propagators and vertices correspond to derivatives of $V(\phi)$.
The first two diagrams contribute to the two terms of $\beta_V$, while the third generates $\beta_Z$.
}
\label{diagrams}
\end{center}
\end{figure}

The Landau-Ginzburg description of the {\tt Blume-Capel} class consists of an action
\begin{equation}\label{action}
S[\phi] = \int {\rm d}^d x \left\{{\textstyle{\frac{1}{2}}}(\partial\phi)^2+V(\phi)\right\}\,,
\end{equation}
in which the potential becomes quintic at criticality.
We renormalized \eqref{action} using minimal subtraction ($\overline{\rm MS}$) of the $\frac{1}{\epsilon}$ poles in $d=\frac{10}{3}-\epsilon$
and we used the results to construct beta functions for the effective potential $V(\phi)$ and a wave function $Z(\phi)$ in a background field approach.
The relevant diagrams for the leading contributions to the flow appear at three loops and are shown in Fig.~\ref{diagrams}
(the next-to-leading contribution is at six loops).

The beta functionals are
\begin{equation}\label{beta-functionals}
\begin{split}
\beta_V &= a \left(V^{(2)}(V^{(4)})^{2}-\frac{9}{2}(V^{(3)})^{2}V^{(4)} \right)\,,\\
\beta_Z &= - b(V^{(5)})^{2}\,,
\end{split}
\end{equation}
where we defined the positive coefficients
\begin{equation}
\begin{split}
& a  = \frac{\Gamma(\frac{1}{2})^{4}\Gamma(\frac{2}{3})}{9(4\pi)^{5}\Gamma(\frac{4}{3})^{2}}\,,
\quad  b = \frac{3\Gamma(\frac{2}{3})^{3}}{40(4\pi)^{5}} \,. 
\end{split}
\end{equation}
We also checked agreement with the beta functionals of \cite{ODwyer:2007brp,Codello:2017hhh},
from which \eqref{beta-functionals} can be evinced by analytically continuing the next-to-leading terms
of the {\it even} models $\phi^{2n}$ to $n=5/2$.
The beta functional $\beta_V$ should be understood as generating function of the beta functions of the couplings of the local operators $\phi^k$,
and, following the discussion of \cite{Codello:2017hhh}, the system \eqref{beta-functionals} captures unambiguously all contributions to the RG flow of all the
relevant operators and the first irrelevant one ($k=1,\dots,5$).

\section{Critical exponents}

Critical properties must be investigated in units of the RG scale $\mu$.
We define the dimensionless potentials
\begin{eqnarray*}\label{dimensionless-potentials}
 v(\varphi) &=& \mu^{-d} V(Z_0^{-1/2} \, \mu^{d/2-1} \, \varphi)\,,
 \\
 z(\varphi) &=& Z_0^{-1} Z(Z_0^{-1/2} \, \mu^{d/2-1} \, \varphi)\,,
\end{eqnarray*}
which include the rescaling of the field by the square root of $Z_0=Z(0)$
in order to have a canonically normalized kinetic term,
and introduce an anomalous dimension $\eta = -\beta_{Z_0}/Z_0$.
For future purpose and for simplifying the result, we further rescale $ v \to \frac{2}{3}(4\pi)^{5/2}\Gamma(\frac{2}{3})^{-3/2}\, v$.
The beta functionals are
\begin{equation}\label{beta-functionals-dimensionless}
\begin{split}
 \beta_v 
  &= -\frac{10}{3} v + \frac{2}{3}\varphi v' + \epsilon \left(v-\frac{1}{2}\varphi v' \right) + \frac{\eta}{2}\varphi v' \\
  &+ \frac{1}{3} v^{(2)}(v^{(4)})^2 -\frac{3}{2} (v^{(3)})^2 v^{(4)}\,,
  \\
  \beta_z 
  &= \frac{2}{3}\varphi z' + \eta \left( z +\frac{1}{2} \varphi z'\right) -\frac{1}{30} (v^{(5)})^2\,.
\end{split}
\end{equation}
The dimensionless wave function satisfies $z(0)=1$ by construction, thus its flow can be used to determine the anomalous dimension
as a function of the dimensionless potential giving $\eta = \frac{1}{30} (v^{(5)}(0))^2$.

The fixed point solutions of $\beta_v=0$ from \eqref{beta-functionals-dimensionless} is a quintic potential
of the form $v(\varphi) = g \varphi^5$ with the constant $g$ being a function of $\epsilon$.
For the comparison with standard perturbation theory, we find convenient to consider $g$ as the critical coupling
which has beta function
\begin{eqnarray*}
 \beta_g &=& -\frac{3}{2}\epsilon g - \frac{153}{4}(5!)^2 g^3\,.
\end{eqnarray*}
The fixed points of $g$ are in one-to-one correspondence with fixed points of \eqref{beta-functionals-dimensionless} in the form  $g \varphi^5$.
There is a complex-conjugate pair of nontrivial purely imaginary solutions
\begin{eqnarray}\label{fixed-point}
 g(\epsilon) &=& \frac{\sqrt{-\epsilon}}{60\sqrt{102}}\,.
\end{eqnarray}
The expansion of critical solutions is thus in semi-odd powers of $\epsilon$ which has been long well known \cite{Nicoll:1974zz}.

The critical exponents can be obtained by linearizing the RG flow \eqref{beta-functionals-dimensionless} around the fixed point solution \eqref{fixed-point}
and diagonalizing its stability matrix. For this we parametrize $v(\varphi)=\sum^5_{i=0} g_i \varphi^i$, thus including all relevant operators and $\varphi^5$.
Within this basis of operators and up to the first order in $\epsilon$, the stability matrix is already diagonal and the scaling operators coincide with $\varphi^i$ for $i=0,\dots,5$.
It is thus convenient to express the critical exponents $\theta_i$ in terms of the operators' anomalous dimensions $\tilde{\gamma}_i$
(we follow the notation of \cite{Codello:2017hhh} in which quantities with tilde are computed with $\overline{\rm MS}$)
\begin{equation}
\begin{split}
 \theta_i &= \frac{10}{3}-\frac{2i}{3}+\epsilon\Bigl(-1+\frac{i}{2}\Bigr) -\tilde{\gamma}_i \\
 \tilde{\gamma}_i &= \frac{\epsilon }{153}\Bigl( \frac{52}{5}i-\frac{139}{12}i^2-\frac{1}{2}i^3+\frac{19}{12}i^4-\delta_{i,5}\Bigr)\,,
\end{split}
\end{equation}
in which we use the determination of the anomalous dimension at the critical point
\begin{equation}
\begin{split}
 \eta &= 2\tilde{\gamma}_1=4 \cdot 5! \cdot g(\epsilon)^2 = -\frac{\epsilon}{765} \,.
\end{split}
\end{equation}
The critical exponents satisfy the scaling relations $\theta_1+\theta_4=\theta_0=d$, $\theta_1=(d+2-\eta)/2$, $\theta_4=(d-2+\eta)/2$ \cite{Codello:2017hhh}.
A comparison of our leading estimate for $\eta$ with the result given in \cite{Gracey:2017okb}
shows some
disagreement, even when taking into account the different conventions.
However, we can provide several further consistency checks of our results (see also the following section).

We give numerical estimates for some notable critical exponents: the anomalous dimension $\eta$,
the exponent $\sigma=\theta_4/\theta_1$, the correlation length exponent $\nu\equiv (\theta_2)^{-1}$,
and a subleading magnetization exponent $\zeta=\theta_3/\theta_1$.
Setting $\epsilon=1/3$ we find
\begin{eqnarray}
\begin{array}{lll}
 \eta = -4.357 \cdot 10^{-4}\,, &\qquad& \sigma =0.2030\,, \\
 \nu = 0.4977\,, &\qquad& \zeta = 0.5596\,.
\end{array}
\end{eqnarray}
We do not estimate the correction-to-scaling exponent $\omega=-\theta_5=3\epsilon$
(which is related to the subleading energy exponent)
because it is expected to receive large corrections from the next-to-leading orders
of the $\epsilon$ expansion.
One interesting property is that the leading quantum/statistical fluctuations drive the correlation length exponent
$
\nu=\frac{1}{2}-\frac{7\epsilon}{1020}
$
to values that are lower than the mean field $\nu_{\rm MF}=1/2$ below the upper critical dimension.
This does not happen to the {\tt Ising} and {\tt Lee-Yang} universality classes.
Whether this property is stable under further corrections requires further study.

\section{CFT data}

We now turn our attention to the characterization of the CFT data of the universality class.
The scaling dimensions of the relevant operators are defined as $\Delta_i=d-\theta_i$.
The case $i=4$ is excluded because the operator $\phi^4$ is a CFT descendant
due to the equations of motion $\partial^2\phi\sim \phi^4$.
Our three-dimensional numerical estimates are
\begin{eqnarray*}
\begin{array}{lllll}
 \Delta_{1} = 0.4998\,, &\quad& \Delta_{2} =0.9908\,, &\quad& \Delta_{3} = 1.5908\,. 
\end{array}
\end{eqnarray*}
The versatility of the functional approach allows for estimates
of some of the OPE coefficients and therefore of the structure constants of the CFT (see \cite{Codello:2017hhh}).
Given the symmetrized fusion rules
$$
 \phi^{(i} \times \phi^{j)} = \sum_k \, C^k{}_{ij} \phi^k+\dots \,,
$$
the $\overline{\rm MS}$ scheme provides the following estimates
\begin{equation}
\begin{split}
  &\tilde C^k{}_{ij} =
  \frac{10}{3}  i(i-1)j(j-1)g(\epsilon)
  \Bigl\{
  648-444(i+j)
  \\
  &\quad
  +78(i^2+j^2)+266ij-37(i^2j+ij^2)
  \\
  &\quad
  +2i^2j^2
  \Bigr\}\delta_{i+j-k,5}
  +2g(\epsilon) \Bigl\{
  i\delta_{i,k}\delta_{j,5}
  +j\delta_{j,k}\delta_{i,5}
  \Bigr\}
  \\
  &\quad
  + 10 g(\epsilon) \delta_{i,5}\delta_{j,5}\delta_{k,5}\,,
\end{split}
\end{equation}
for the $C^k{}_{ij}$, whenever $i+j-k=5$. These expressions
require the use of $g(\epsilon)$ given in \eqref{fixed-point}.
The estimates are unaffected by mixing with higher derivative operators for $i,j,k\leq 5$.

In \cite{Codello:2017qek} the {\tt Blume-Capel} universality class was considered in $d=\frac{10}{3}-\epsilon$
on purely CFT grounds using a method that allows to build a conformal theory out of the free theory's data.
The pure CFT construction does not yet determine all CFT data, but it gives some quantities to compare with.
The following ratio is independent of the FP coupling and agrees with the same quantity as given in \cite{Codello:2017qek}
$$
\frac{\tilde{\gamma}_{2}}{\tilde{\gamma}_{1}}=\frac{2-\theta_{2}}{\frac{1}{2}\eta}=\frac{-\frac{7\epsilon}{255}+O(\epsilon^{2})}{-\frac{1}{2}\frac{\epsilon}{765}+O(\epsilon^{2})}=42+O(\epsilon)\,,
$$
in which we restored the order of approximation to clarify which is the next-to-leading contribution.
Additionally, we can compare some ratios involving the $\tilde{C}^k{}_{ij}$ with the corresponding structure functions of the CFT three point functions $C_{ijk}$
which are related to the OPE coefficients $C^k{}_{ij}$.
The ones that can be compared with \cite{Codello:2017qek} are
\begin{eqnarray} \label{univ_ope_bc}
 &  \frac{\tilde C^1{}_{15}}{\sqrt{\tilde\gamma_1}} = 4\sqrt{15}+O(\epsilon)\,, \nn
 \quad
 \frac{\tilde C^1{}_{24}}{\sqrt{\tilde\gamma_1}}
 = 32\sqrt{15}+O(\epsilon)\,,
 \\
 & \frac{\tilde C^1{}_{33}}{\sqrt{\tilde\gamma_1}}
 = -108\sqrt{15}+O(\epsilon)\,, 
\end{eqnarray}
and they agree exactly. In fact, given our estimate of $\tilde\gamma_1$ in terms of $\epsilon$,
we can use the results
of \cite{Codello:2017qek} to find 
the leading $\epsilon$ dependence of a family of structure constants which includes \eqref{univ_ope_bc} 
\begin{equation}
C^1{}_{kl} = \frac{k!l!}{\left(\!\frac{4+l-k}{2}\!\right)!\left(\!\frac{4+k-l}{2}\!\right)!\left(\!\frac{k+l-4}{2}\!\right)!}\frac{2\sqrt{-\epsilon}}{(k\!-\!l)^2\!-\!1}\sqrt{\frac{6}{17}}\,,
\end{equation}
in which $ \left|k-l\right|\leq 4$ and $k+l\geq 4$ with $k$ and $l$ being both even or odd, as well as other constants with a leading $\epsilon$ contribution
\begin{equation}
C^1{}_{14} = \frac{9\epsilon}{17}\,, \quad
C^1{}_{16} = -\frac{20\epsilon}{51}\,, \quad
C^1{}_{18} = -\frac{7\epsilon}{51}\,, 
\end{equation}
and one constant with leading $(-\epsilon)^{\frac{3}{2}}$ contribution
\begin{equation}
C^1{}_{11} = \left(\frac{3}{34}\right)^{\frac{3}{2}}(-\epsilon)^{\frac{3}{2}}\,.
\end{equation}
Evaluating these results at $\epsilon=\frac{1}{3}$ provides a first numerical estimate for the structure constants of the three-dimensional
{\tt Blume-Capel} class.

\section{Other multicritical models}

It is not difficult to generalize the results presented so far to the entire family of odd multicritical models $\phi^{2n+1}$.
These models are interesting in their own respect but, besides the {\tt Lee-Yang} ($n=1$) and {\tt Blume-Capel} ($n=2$) universality classes,
they have upper critical dimension $d_c<3$ and therefore only physical $d=2$ realizations,
apart from possible \emph{fractal} realizations.
Their upper critical dimension is
\begin{equation}
 \begin{split}
  d_c &= 2+\frac{4}{2n-1}\,,
 \end{split}
\end{equation}
and they can be renormalized starting from the action \eqref{action} and subtracting the poles in $d=d_c-\epsilon$.
The diagrams involved in this subtraction have $2n-1$ loops and generalize those of Fig.~\ref{diagrams}.

All the steps leading to the scaling analysis of the $n=2$ case can be followed through and are mostly unchanged,
including the definition of dimensionless potential $v(\varphi)$.
The convenient rescaling as a function of $n$ is
\be \label{rescaling}
v \to \frac{8}{2n-1}\, c^{-\frac{2n-1}{2}}\, v \quad {\rm with}\quad c = \frac{1}{4\pi}\frac{\Gamma(\delta_n)}{\pi^{\delta_n}}\,,
\ee
and replaces the one of the previous sections. We denoted with $\delta_n=\frac{2}{2n-1}$ the canonical dimension
of the field $\phi$ at $d_c$.
The general dimensionless beta functionals in terms of the label $n$ are given as
\begin{equation}\label{bv_general}
\begin{split}
\beta_v &= -d v \!+\! \frac{d\!-\!2\!+\!\eta}{2} \varphi  v'\\
&  -\frac{1}{3}\,\Gamma(1\!+\!\delta_n)\hspace{-0.045\textwidth}\sum_{\footnotesize \begin{array}{c}\raisebox{-3pt}{$r\!+\!s\!+\!t\!=\!2n\!+\!1$}\end{array}}\hspace{-0.04\textwidth}\frac{B^n_{rst}}{r!s!t!} \;v^{(r+s)}\,v^{(s+t)}\,v^{(t+r)}\,, \\[7pt]
\beta_z &= \eta z \!+\! \frac{d\!-\!2\!+\!\eta}{2} \varphi  z'\! -\!  \frac{4}{(2n\!+\!1)!} (v^{(2n\!+\!1)})^2\,, \\
\end{split}
\end{equation}
where $r,s,t \in \mathbb{N}$. We defined the coefficients
\begin{equation}
B^n_{rst}=A^n_{r,st}\cdot A^n_{s,tr}\cdot A^n_{t,rs}\,, \quad A^n_{r,st}=\frac{
\Gamma\left(\frac{(s+t-r)}{2}\delta_n\right)
}{
\Gamma\left(r\delta_n\right)
}\,.
\end{equation}
One can check that this formula reproduces \eqref{beta-functionals-dimensionless} for $n=2$ and that of the leading contribution to the {\tt Lee-Yang}
universality class given in \cite{Codello:2017hhh}.

The anomalous dimension $\eta$ can be read off imposing $z(0)=1$ in \eqref{bv_general} to obtain
\begin{equation} \label{eta_general}
\eta = 4(2n+1)!g^2\,,
\end{equation}
where $g$ comes from the critical potential $v(\varphi)=g\varphi^{2n+1}$.
This can be used in $\beta_v$ to find the interacting fixed point value
{\setlength\arraycolsep{2pt}
\begin{eqnarray}\label{general-fixed-point}
 g(\epsilon) &=&   \bigg[(2n+1)!^3\,\Gamma(\delta_n)\sum_{{\footnotesize r,s,t}}\frac{B^n_{rst}}{(r!s!t!)^2} \\
&&  - 3(4n^2-1)\,(2n+1)!\bigg]^{-\frac{1}{2}}\frac{2n-1}{6}\sqrt{-\epsilon}\,, \nn
\end{eqnarray}}%
where the summation runs over the same indices as \eqref{bv_general}.
The fixed point can be used in the expression for $\eta$ to find its leading contribution.

From the Taylor expansion of the RG flow of \eqref{bv_general} at \eqref{general-fixed-point} we find the anomalous dimensions 
{\setlength\arraycolsep{2pt}
\begin{eqnarray} \label{gamma_general}
\tilde\gamma_i &=& \! (2n+1)!\,\bigg[2i + 4(2n+1)\, \delta^{2n+1}_i 
\\
&&\hspace{-11pt} -(2n\!+\!1)!i!\,\Gamma(1\!+\!\delta_n) \sum_{{\footnotesize r,s,t}}\frac{B^n_{rst}}{r!s!^2t!^2} \;\frac{1}{(i\!-\!s\!-\!t)!}\bigg]g(\epsilon)^2  \nn
\end{eqnarray}}%
and the estimates of some of the OPE coefficients
\begin{eqnarray} \label{ope_general}
\tilde C^k{}_{ij} 
&=&  -\Gamma(1\!+\!\delta_n)\sum_{{\footnotesize r,s,t}}\frac{B^n_{rst}}{r!s!t!^2} \frac{i!j!(2n\!+\!1)! \,g(\epsilon)}{(j\!-\!s\!-\!t)!(i\!+\!s\!-\!2n\!-\!1)!} \nonumber \\
&& \hspace{-29pt} + 2(2n\!+\!1)!\left(\!i\delta^{2n\!+\!1}_j \!+\! j\delta^{2n\!+\!1}_i\!+\!(2n\!+\!1)\,\delta^{2n\!+\!1}_i\delta^{2n\!+\!1}_j\!\right)\!g(\epsilon) \nn
\end{eqnarray}
when $i\!+\!j\!-\!k\!=\!2n\!+\!1$.
Notice that the latter vanish in the free-theory limit
and that $\overline{\rm MS}$ gives access only to the ``massless'' OPE coefficients according to \cite{Codello:2017hhh}. 

Let us show explicitly some specific quantities.
We have that $\tilde\gamma_1$ can be shown to be $\eta/2$ appearing in \eqref{eta_general}.
The anomalous scaling of the mass
$\tilde\gamma_2$ gets contributions only from $s=t=1$ in \eqref{gamma_general}
{\setlength\arraycolsep{2pt}
\begin{eqnarray} \label{gamma2_general}
\tilde\gamma_2
&=& 4(2n\!+\!1)!\,(2n-1)\,\frac{2n+3}{2n-3}\, g(\epsilon)^2\,.
\end{eqnarray}}%
The OPE coefficients with index $k=1$ are 
\begin{eqnarray} \label{ope1_general}
\tilde C^1{}_{i, 2n+2-i}
&=& \frac{2i(2n\!+\!2\!-\!i)(2n\!+\!1)!(2n-1)}{(2n-2i+3)(2n-2i+1)}\,g(\epsilon)\,,\nn
\end{eqnarray}
where we neglected the marginal cases in which either $i$ or $j$ equal $2n+1$.
All the above explicit results reproduce formulas given in \cite{Codello:2017qek}, provided that one restores the factor rescaled away by \eqref{rescaling}
as well as the factorials,
i.e.\ one makes the replacement $ g\to \frac{2n-1}{8} c^{\frac{2n-1}{2}} \frac{g}{(2n+1)!} $.
%

\smallskip
\smallskip

\section{Discussion and Outlook}

In this paper we have reported a detailed
analysis of scaling and conformal properties
of the {\tt Blume-Capel} universality class in the $\epsilon$ expansion.
Our results are interesting for two main reasons: This universality class has been
mostly ignored up to now (with some exceptions \cite{Zambelli:2016cbw,Codello:2017qek,Gracey:2017okb}),
even though it is nontrivial (and nonunitary) in dimension three.
Its upper critical dimension is fractional and just above three,
which presumably makes the $\epsilon$ expansion more reliable.

We have given some numerical estimates of universal
quantities in three dimensions where $\epsilon=\frac{1}{3}$, in the hope that a numerical simulation
might confirm our findings.
Based on the Euclidean/Lorenzian duality arguments of CFT, a candidate Lorenzian lattice model that might exhibit
this universal behavior is the spin one Blume-Capel model (hence the name) on a two-dimensional grid (thus in $2+1$ dimensions)
and criticality should be achieved by tuning the magnetic field to a purely imaginary value.
The spin one Hamiltonian should give enough local degrees of freedom to probe a
tricritical phase which should occur at imaginary magnetic field because of nonunitarity.

To promote the importance of the {\tt Blume-Capel} universality class,
it would be interesting to understand if its Landau-Ginzburg form corresponds to a minimal CFT in
two dimensions. According to \cite{vonGehlen:1994rp,Zambelli:2016cbw} and following the logic of the previous paragraph,
this minimal model could either be ${\cal M}_{2,7}$ or ${\cal M}_{2,9}$ (see also \cite{Mossa:2007fx} in relation to the Yang-Lee edge singularity).
While there is no definite answer yet, we believe that it could be achieved using, for example,
the methods of \cite{Milsted:2017csn}.

We have also studied the whole family of multicritical odd models $\phi^{2n+1}$.
The general results compare well with, and generalize, an analysis based on CFT methods of the same models,
thus strengthening their status as conformal theories \cite{Codello:2017qek}. However for all models $n>2$
the upper critical dimension is smaller than three, implying that they have only two-dimensional physical realizations
(besides possible realizations on \emph{fractals})
in which they could be interpreted as multicritical generalizations of the {\tt Lee-Yang} universality class.
It would be an outstanding theoretical achievement to understand which conformal theories these models correspond to
in the two dimensional limit.

\paragraph*{Acknowledgments}
O.Z.\ acknowledges support by the DFG under grant No.~Gi328/7-1.
A.C.\ and O.Z.\ are grateful to INFN Bologna for hospitality and support.
We thank J.~A.~Gracey for feedback on the draft.



\begin{thebibliography}{99}


\bibitem{Pelissetto:2000ek} 
  A.~Pelissetto and E.~Vicari,
  Phys.\ Rept.\  {\bf 368}, 549 (2002)
  [cond-mat/0012164].
  
\bibitem{Belavin:1984vu} 
  A.~A.~Belavin, A.~M.~Polyakov and A.~B.~Zamolodchikov,
  Nucl.\ Phys.\ B {\bf 241}, 333 (1984).

\bibitem{Brezin:1972fc} 
  E.~Brezin, D.~J.~Wallace and K.~G.~Wilson,
  Phys.\ Rev.\ Lett.\  {\bf 29}, 591 (1972).

\bibitem{Fisher:1978pf} 
  M.~E.~Fisher,
  Phys.\ Rev.\ Lett.\  {\bf 40}, 1610 (1978).
  
\bibitem{Nicoll:1974zz} 
  J.~F.~Nicoll, T.~S.~Chang and H.~E.~Stanley,
  Phys.\ Rev.\ Lett.\  {\bf 33}, 540 (1974).
  
\bibitem{vonGehlen:1994rp} 
  G.~von Gehlen,
  Int.\ J.\ Mod.\ Phys.\ B, 08, 3507 (1994)
  [hep-th/9402143];
  Nucl.\ Phys.\ B {\bf 330}, 741 (1990).
  
\bibitem{Zambelli:2016cbw} 
  L.~Zambelli and O.~Zanusso,
  Phys.\ Rev.\ D {\bf 95}, no.\ 8, 085001 (2017)
  [arXiv:1612.08739 [hep-th]].

\bibitem{ElShowk:2012ht} 
  S.~El-Showk, M.~F.~Paulos, D.~Poland, S.~Rychkov, D.~Simmons-Duffin and A.~Vichi,
  Phys.\ Rev.\ D {\bf 86}, 025022 (2012)
  [arXiv:1203.6064 [hep-th]].
  
\bibitem{Gliozzi:2016ysv} 
  F.~Gliozzi, A.~Guerrieri, A.~C.~Petkou and C.~Wen,
  Phys.\ Rev.\ Lett.\  {\bf 118}, no.\ 6, 061601 (2017)
  [arXiv:1611.10344 [hep-th]].
  
\bibitem{Alday:2016njk} 
  L.~F.~Alday,
  Phys.\ Rev.\ Lett.\  {\bf 119}, no.\ 11, 111601 (2017)
  [arXiv:1611.01500 [hep-th]].
  
\bibitem{Rychkov:2015naa} 
  S.~Rychkov and Z.~M.~Tan,
  J.\ Phys.\ A {\bf 48}, no.\ 29, 29FT01 (2015)
  [arXiv:1505.00963 [hep-th]].
  
\bibitem{Basu:2015gpa} 
  P.~Basu and C.~Krishnan,
  JHEP {\bf 1511}, 040 (2015)
  [arXiv:1506.06616 [hep-th]].
  
\bibitem{Nii:2016lpa} 
  K.~Nii,
  JHEP {\bf 1607}, 107 (2016)
  [arXiv:1605.08868 [hep-th]].
  
\bibitem{Codello:2017qek}
  A.~Codello, M.~Safari, G.~P.~Vacca and O.~Zanusso,
  JHEP {\bf 1704} (2017) 127
  [arXiv:1703.04830 [hep-th]].

\bibitem{ODwyer:2007brp}
  J.~O'Dwyer and H.~Osborn,
  Annals Phys.\  {\bf 323}, 1859 (2008)
  [arXiv:0708.2697 [hep-th]].

\bibitem{Codello:2017hhh} 
  A.~Codello, M.~Safari, G.~P.~Vacca and O.~Zanusso,
  arXiv:1705.05558 [hep-th].

\bibitem{BC}
 M.~Blume, Phys.\ Rev.\ {\bf 141}, pp.~517 (1966);
 H.~W.~Capel, Physica 32 pp.~966 (1966).

\bibitem{Mossa:2007fx} 
  A.~Mossa and G.~Mussardo,
  J.\ Stat.\ Mech.\  {\bf 0803}, P03010 (2008)
  [arXiv:0710.0991 [hep-th]].

\bibitem{Castro-Alvaredo:2017udm} 
  O.~A.~Castro-Alvaredo, B.~Doyon and F.~Ravanini,
  arXiv:1706.01871 [hep-th].
  
\bibitem{Gracey:2017okb} 
  J.~A.~Gracey,
  arXiv:1703.09685 [hep-th].
  
\bibitem{Milsted:2017csn} 
  A.~Milsted and G.~Vidal,
  arXiv:1706.01436 [cond-mat.str-el].
  
\end{thebibliography}
\end{document}